\documentstyle[12pt]{article}
\begin{document}
\def\baselinestretch{1.5}
\newcommand{\lb}{\label}
\newcommand{\be}{\begin{equation}}
\newcommand{\ee}{\end{equation}}
\newcommand{\fr}{\frac}
\newcommand{\D}{\delta}
\newcommand{\sig}{\sigma}
\newcommand{\del}{\partial}
\newcommand{\wv}{\wedge}
\newcommand{\al}{\alpha}
\newcommand{\la}{\lambda}
\newcommand{\La}{\Lambda}
\newcommand{\ep}{\epsilon}
\newcommand{\pr}{\prime}
\newcommand{\ti}{\tilde}
\newcommand{\om}{\omega}
\newcommand{\OO}{\Omega_{BRST}}
\newcommand{\bi}{\bibitem}

\rightline{IC/92/428}

\rightline{December, 1992}

\vspace{2cm}

\begin{center}

{\large
A GENERAL SOLUTION OF THE MASTER EQUATION

\vspace{1cm}

FOR A CLASS OF FIRST ORDER SYSTEMS}

\vspace{2cm}
{\normalsize
\"{O}mer F. DAYI}

{\small \it
University of Istanbul,
Faculty of Science,  \\
Department of Physics,
Vezneciler, 34459 Istanbul, Turkey, \\
and \\
International Centre for Theoretical Physics,\\
 P.O.Box 586, 34100-Trieste, Italy}

\vspace{2.5cm}

\end{center}

{\small

Inspired by the  formulation of the Batalin-Vilkovisky method of
quantization in terms of  ``odd time'',
we show that for a class of gauge theories which are first order
in the derivatives, the kinetic term is bilinear in
the fields, and the interaction part satisfies some properties, it is
possible to
give the solution of  the master equation in a very simple way.
To clarify the general
procedure we discuss its application
to Yang-Mills theory, massive
(abelian) theory in the Stueckelberg formalism, relativistic particle
and to the self-interacting antisymmetric tensor field.}

\pagebreak

The formulation of the Batalin-Vilkovisky (BV) method of
quantization\cite{BV} in terms of a Grassmann odd parameter
behaving as time (``odd time") was helpful to
derive systematically the ``ad hoc" definitions of
Batalin and Vilkovisky\cite{o1}. In Ref.\cite{o1}
existence of an
appropriate Lagrangian for the odd time formulation
(``odd time Lagrangian")
was assumed. Although it is possible
to write an odd time Lagrangian by using the Hamiltonian formalism
of the BV method\cite{hbv}, it does not give new
hints about finding a solution of the master equation
of the Lagrangian formalism.
In the Hamiltonian formalism a general solution of
the master equation is available in terms of the
Becchi-Rouet-Stora-Tyutin (BRST)-charge
which gives a vanishing generalized Poisson bracket with
itself.

Recently, the extended form method developed in Ref.\cite{cc}
was utilized to write actions  for topological quantum field
theories\cite{GMDM} which lead to solution of the master
equation\cite{ikemori}.

Inspired by these, we show that for a class of gauge theories
which are first order in the derivatives, the kinetic
part is bilinear in the fields (variables), and the
interaction terms possessing some properties,
it is possible to write an action which simply
leads to the  solution of the master equation in the
minimal ghost sector. In fact, the actions of Ref.\cite{ikemori}
can be obtained from this general method.

First we recall the basic concepts of the odd time formulation
of the BV method and discuss the  related Lagrangian.
This discussion is not essential for the rest of the
paper, but it is useful to understand how the solution of
the master equation is inspired.

We give the rules of constructing an action in terms
of the minimal ghost fields, the antifields and the
initial gauge theory action satisfying some conditions, and
prove that it is the desired solution of the master equation. We
illustrate the method by its application to Yang-Mills theory, massive
(abelian) theory in Stueckelberg formalism, the relativistic particle
and the self-interacting antisymmetric tensor field.

When we deal with a gauge theory we can introduce
the  odd time $\tau_0$
(a parameter possessing odd Grassmann parity),
such that the change of a function $f$ by
the BRST-charge $\OO $,
is written symbolically as
\[
\OO f=\fr{\del f(\tau_0)}{\del \tau_0}.
\]
We assume that there exists an odd time
Lagrangian
$ L( \Phi(\tau_0), \dot{\Phi}(\tau_0) ) $,
which carries  information about
the BRST transformations. $\Phi$ includes the original
fields of the original gauge theory, and the related
ghost fields;
      $ \dot{\Phi}(\tau_0) \equiv \del \Phi (\tau_0) / \del \tau_0$.
The ``odd time canonical momentum" which results from this Lagrangian is
\be
\Pi (\tau_0)=\fr{\del L (\Phi (\tau_0) ,\dot{\Phi}(\tau_0))}
{\del \dot{\Phi}(\tau_0) }.
\ee

On the cotangent bundle of a supermanifold an odd canonical
two form is known to exist when
the cotangent bundle has an equal number of odd and even coordinates
\cite{leites}. Thus we can define
an ``odd Poisson bracket'' (antibracket)
\be
\label{abrac}
(f,g) \equiv
\fr{\del_r f}{\del \Phi}
\fr{\del_l g}{\del \Pi} -
\fr{\del_r f}{\del \Pi}
\fr{\del_l g}{\del \Phi},
\ee
where $\del_r$ and $\del_l$ indicate the right and the left derivatives.
In this phase space odd time evolution is given by the Grassmann-even
Hamiltonian $S$:
\be
\fr{\del f}{\del \tau_0}=(S,f).
\ee
Thus $S$ must satisfy
\be
\label{master}
\fr{\del S}{\del \tau_0} =(S,S) =0.
\ee
This is the master equation of Batalin and Vilkovisky.

The easist way of defining an odd time Lagrangian is to take it
independent of the velocities. Then
\[
S(\Phi )= -L(\Phi ),
\]
and the master equation is automatically satisfied. Of course, we should
give the conditions to construct $L(\Phi )$,
such that $S(\Phi )$ is the
action of the BV method of quantization.

To gather the original fields and the ghosts one can
extend the ordinary
differential forms to include also the ghost number. This
can be achieved by
generalizing the exterior derivative as \cite{cc}
\be
\lb{dtil}
d \rightarrow  \ti{d} \equiv d + \delta ,
\ee
where $\delta $ denotes the BRST transformation.
In order to utilize  this generalization of $d$ to find the
solution of the master equation we follow the following
procedure.

$i )$ If the original gauge theory is not already first order
in $d$ and the terms containing $d$ are not bilinear in fields,
one should find an equivalent formulation of it possessing these
properties.

$i i )$  The minimal ghost content of the theory should be found by
analyzing the related gauge invariance and the proper solution
condition of Batalin and Vilkovisky. Generalize the
original fields to include also the ghosts and antifields which
possess the same grading  with the  original ones
in terms of $\tilde{d}$. Now, substitute the original fields
$\phi$, with the generalized ones
$\Phi \equiv (\ti{A},  \ti{B} ) $, in the Lagrangian.
The resulting action is the one which can be used in the BV method
if it is in the form
\be
\lb{ef}
S(\Phi) =\ti{B} d\ti{A}  +\al \ti{B}
\ti{B}  +\beta \ti{A}  \ti{A}  + \gamma \ti{A} \ti{A} \ti{B} ,
\ee
where $\al$ or $\beta$ vanishes and the BRST transformation
of the fields
\be
\lb{tr}
\delta \ti{A}  =\fr{\del_l S}{\del \ti{B} },\
\delta \ti{B}  =-\fr{\del_l S}{\del \ti{A} },
\ee
can be written in terms of
$\ti{D} = d+\ti{A} $ and the related curvature $\ti{F}$,
when $\beta =0$ as
\be
\lb{bz}
\delta \ti{A}  = \ti{F}  -\ti{B}  ,\  \delta \ti{B}  = -\ti{D} \ti{B}  ,
\ee
and when $\alpha =0$ as
\be
\lb{az}
\delta \ti{A}  = \ti{F}   ,\   \delta \ti{B}  = -\ti{D} \ti{B}  +\ti{A}  .
\ee
In (\ref{ef}) multiplication is defined such that $S$ is a
scalar possessing zero ghost number. In (\ref{tr})-(\ref{az})
the components of the right hand side are restricted to
possess the same grading and one more ghost number of the
components of the left hand side.

It is possible to choose the signs of the components of
$\ti{A} $ and $\ti{B} $ such that
\[
\delta_1S \equiv \fr{\del_r S}{\del \ti{B} } \delta \ti{B} = \fr{1}{2}
(S,S),
\]
\[
\delta_2S \equiv \fr{\del_r S}{\del \ti{A} } \delta \ti{A} = \fr{1}{2}
(S,S),
\]
where $(S,S)$ is the appropriate master equation. By  taking the
derivative of these with respect to $\ti{A} $ and $\ti{B} $, one can
show that $S$ satisfies the master equation ( $(S,S)=constant
\neq 0 $ would lead to the non-consistency of equations
of motion) if
\[
\delta^2 \ti{A}  =0,\   \delta^2 \ti{B}  =0 ,
\]
where for an arbitrary functional $f$, BRST transformation is
defined as
\[
\delta f =
\fr{\del_r f}{\del \ti{A} } \delta \ti{A}  +
\fr{\del_r f}{\del \ti{B} } \delta \ti{B}  .
\]

In the case given in (\ref{bz}), formally we have
\[
\begin{array}{lll}
\delta^2 \ti{A}   & = &  \ti{D} \cdot  (\ti{F} -\ti{B} )+\ti{D}
\ti{B}  , \\
\delta^2\ti{B}  & = & -\ti{D}  \cdot  \ti{D} \ti{B}
+ (\ti{F} -\ti{B} )\cdot  \ti{B} .
\end{array}
\]
These vanish due to the Bianchi identities $\ti{D} \cdot  \ti{F} =0$,
the definition of the curvature $\ti{F} =\ti{D} \cdot  \ti{D} $, and
because when the related conditions permit that $\ti{B} ^2$ exists
we have $\ti{B} \cdot  \ti{B}
=\ti{B} _i\ti{B} _j - (-1)^{\ep (\ti{B} _i) ep (\ti{B} _j)}
\ti{B} _j \ti{B} _i =0 $, where $\ep$ indicates the Grassmann parity.

When we deal with the case given in (\ref{az})
\[
\begin{array}{lll}
\delta^2 \ti{A}   & = &  \ti{D} \cdot  \ti{F} , \\
\delta^2\ti{B}  & = & -\ti{D}  \cdot
(- \ti{D} \ti{B} +\ti{A} ) - \ti{F} \ti{B}  + \ti{F}       ,
\end{array}
\]
which vanish due to the Bianchi identities and the definition
of curvature.

In the case $\ti{A} =\ti{B} $ (Chern-Simons type) we have both
$\alpha =0$ and $\beta =0$ in (\ref{ef}) and $\delta \ti{A} =\ti{F} $,
so that $\delta^2 \ti{A} =0$ follows from the Bianchi identities.

By construction $S(\Phi )$ possesses the correct classical
limit. Moreover, $\Phi$ is found
by using the proper solution condition, so that it includes
all the fields of the minimal sector and
because of the form of $S$ (\ref{ef}),
\[
rank\  \left| \fr{\del^2 S}{\del \Phi \del \Phi } \right| = N,
\]
where $N$ is the number of the components of
$\ti{A} $ or $\ti{B} $. Hence, we conclude that  $S(\Phi)$
is the  desired action.
It seems that one could relax the conditions on the interaction part
of the action (\ref{ef}), but a general proof of $(S,S)=0$ is lacking.

To clarify the procedure outlined above, let us  see some applications
of it.

\vspace{.2in}

{\bf 1) Yang-Mills Theory}

\vspace{.2in}

It is defined in terms of the second order action (we suppress
$Tr$)
\be
\lb{ym}
L_0= \fr{-1}{2}\int d^4x\   F_{\mu \nu}  F^{\mu \nu},
\ee
where
$F=d \wedge A + A \wedge A$. The theory given by
\be
\lb{eym}
L=\fr{-1}{2}\int d^4x\   ( B _{\mu \nu}   F^{\mu \nu}
-\fr{1}{2} B_{\mu \nu} B^{\mu \nu} ),
\ee
is equivalent to  (\ref{ym}) on mass-shell, and moreover it
is first order
in $d$. (\ref{eym}) is invariant under the
infinitesimal gauge transformations
\[
\D A_\mu =D_\mu \La\ ,\  \D B_{\mu \nu}= [ B_{\mu \nu}, \La ],
\]
where $D=d + [A,\ ]$ is the covariant derivative. They are irreducible, so
that for the covariant quantization we
need only (in the minimal sector) the ghost field
$\eta$, which possesses ghost number 1.

Generalize the fields of (\ref{eym})
according to (\ref{dtil})  as
\[
A  \rightarrow \ti{A}, \ B \rightarrow \ti{B},
\]
so that one obtains $(\del_\mu A_\nu -\del_\nu A_\mu
\rightarrow 2d\ti{A} )$
\be
S=\fr{-1}{2}\int d^4x\   [\ti{B} (2d\ti{A}
+\ti{A}\ti{A}) -\fr{1}{2} \ti{B}\ti{B} ],
\ee
which is defined to possess 0 ghost number.
Grading of the extended forms $\ti{A}$ and $\ti{B}$, respectively,
are 1 and 2, and their first components
are $A_\mu$ and $B_{\mu \nu}$.

By using the fact that
\[
N_{gh}(\phi) + N_{gh}(\phi^\star ) =-1,
\]
where $N_{gh}$ denotes the ghost number,
we write the generalized fields as
\[
\begin{array}{lcl}
\ti{A} & =  &  A_{(1+0)}+\eta_{(0+1)} - B_{(2-1)}^\star , \\
\ti{B} & = & B_{(2+0)}+ A^\star_{(3-1)}+ \eta^\star_{(4-2)},
\end{array}
\]
where the first number in parenthesis is the order of d-forms
and the second is the ghost number. Here $``^\star"$ indicates
the antifields as well as the Hodge-map. Substitution of these
in (\ref{eym}) and using the property of the multiplication
that the product is different
from zero only when its ghost number vanishes, we
get
\be
\lb{syme}
S=-\int d^4x\ (\fr{1}{2}B_{\mu \nu}F^{\mu \nu} -
B^{\mu\nu} [\eta , B_{\mu\nu}^\star ]
+A_\mu^\star D^\mu \eta +
\fr{1}{2}\eta^\star [\eta ,\eta ]
-\fr{1}{4} B_{\mu\nu} B^{\mu\nu}).
\ee
We may perform a partial gauge fixing $B^\star =0$, and then
use the equations of motion with respect to $B_{\mu\nu}$ to obtain
\[
S\rightarrow \ti{S}=- \int d^4x\   (\fr{1}{2}F_{\mu \nu}  F^{\mu \nu}
+A^\star_\mu    D^\mu \eta
+\fr{1}{2}\eta^\star [\eta ,\eta] ),
\]
which is the minimal solution of the master equation for
Yang-Mills theory.

\vspace{.2in}

{\bf 2) Massive Abelian Theory in Stueckelberg Formalism }

\vspace{.2in}

It is defined in terms of the second order Lagrangian
\be
\lb{mts}
L_{0}=-\int d^4x\   [\fr{1}{2} F_{\mu \nu}
F^{\mu \nu} +m^2 (A_\mu -m^{-1}\del_\mu v)
(A^\mu -m^{-1}\del^\mu v)],
\ee
where $F=d \wv A$.
The action linear in $d$ and possessing a kinetic term bilinear in
the fields,
\be
\lb{emts}
L=-\int d^4x\   [\fr{1}{2} B_{\mu \nu} (d\wv A)^{\mu \nu}
-\fr{1}{4}B_{\mu \nu} B^{\mu \nu}
+m (A_\mu -m^{-1}\del_\mu v) K^\mu   -\fr{1}{2}K_\mu K^\mu ],
\ee
is equivalent to (\ref{mts}) on mass-shell.
It is invariant under the gauge transformations
\[
\begin{array}{ll}
\D A_\mu = \del_\mu \La , &  \D B_{\mu \nu } =0, \\
\D v =m\La , &   \D K_\mu =0,
\end{array}
\]
which are irreducible, so that in the minimal sector
there is only one ghost: $\eta$.
By substituting the original fields with the
generalized ones  we get
\be
S=\int d^4x\   [- \ti{B} d\ti{A} + \ti{B}\ti{B}
+m(\ti{A} -m^{-1}d\ti{v})\ti{K} +\fr{1}{2}\ti{K}\ti{K}].
\ee
The generalized fields are
\[
\begin{array}{lcl}
\ti{A} & =  &  A_{(1+0)}+\eta_{(0+1)} - B_{(2-1)}^\star , \\
\ti{B} & = & B_{(2+0)}+ A^\star_{(3-1)}+ \eta^\star_{(4-2)}, \\
\ti{v} & =  & v_{(0+0)}- K^\star_{(1-1)}, \\
\ti{K} & = & K_{(3+0)}+ v^\star_{(4-1)}.
\end{array}
\]
By respecting the rules of multiplication one finds in components
\[
S= L -\int d^4x\   [ A^\star_\mu \del^\mu \eta  -m\eta v^\star ].
\]
We can eliminate $B$ and $K$ by using their equations of motion to
obtain
\[
S\rightarrow \ti{S}=\int d^4x\   [
-\fr{1}{2} F^2_{\mu \nu} - m^2 (A_\mu -m^{-1}\del_\mu v)^2
- A_\mu^\star\del^\mu \eta  +m\eta v^\star ].
\]
Indeed, this is the minimal solution of the master equation
for the theory given by (\ref{mts}).

\vspace{.2in}

{\bf 3) Relativistic Particle}

\vspace{.2in}

In terms of the canonical variables
satisfying the Poisson bracket relation $\{ p_\mu ,q^\nu \} =\D^\nu_\mu$,
the relativistic particle is given by
\be
\lb{rp}
L_0 = \int   (p \cdot dq -\fr{1}{2}e p \cdot p ),
\ee
where $dq^\mu = \del_t q^\mu  dt$.
A variable possesses two different gradings: one of them is due
to the one dimensional manifold of $t$ and the other one is
related to the  space-time manifold.

(\ref{rp}) is invariant under
\[
\D q^\mu =p^\mu \La ,\   \D p =0 , \    \D e =\del_t \La .
\]

Now, we generalize the fields as
\[
{[q, e]} \rightarrow  \ti{q};\  p \rightarrow \ti{p}.
\]
$q$ and $e$ are treated on the same footing due to the fact
that there is no $de$ term in (\ref{rp}).
Hence the odd time Hamiltonian is
\be
\lb{rpe}
S=\int [\ti{p} d\ti{q}
-\fr{1}{2}\ti{q}\ti{p}^2],
\ee
where
\[
\begin{array}{lcl}
\ti{q} & =  &  q^\mu_{(1+0+0)}+e_{(0+1+0)}  +\eta_{(0+0+1)}-
p^{\star \mu}_{(1+1-1)},  \\
\ti{p} & =  &  p_{\mu (d-1+0+0)} +q^\star_{\mu (d-1+1-1)}
+e^\star_{(d+0-1)} + \eta^\star_{(d+1-2)}.
\end{array}
\]
The numbers in the parenthesis
indicate, respectively, grading due to space-time,
grading due to 1-dimensional manifold and ghost number.

By calculating the product in components one can show that
\[
S=\int dt\  [p\cdot \del_t q +e^\star \del_t \eta -\fr{1}{2}
e p^2 + q^\star \cdot p \eta ],
\]
which is the minimal solution of the master equation
for the relativistic particle.

\vspace{.2in}

{\bf 4) The Self-interacting Antisymmetric Tensor Field}

\vspace{.2in}

The action\cite{ft} (we suppress $Tr$ again)
\be
\lb{si}
L_0 =-\int d^4x\   [B_{\mu \nu}(d\wv A+A\wv A)^{\mu \nu}
- \fr{1}{2} A_\mu A^\mu ],
\ee
is invariant under the transformations
\[
\D B_{\mu \nu } =\ep_{\mu \nu \rho \sigma }D^\rho \La^\sigma ,\
\D A_\mu =0 ,
\]
and is analysed in terms of the BRST methods in Ref.\cite{af}.
If we set $\La_\mu = D_\mu \al$, the gauge transformation
vanishes on shell $\D B|_{F=0}=0$. This is a first-stage
reducible theory, hence we need to introduce the ghost fields
\[
C_0^\mu , \ C_1 ;\  N_{gh}(C_0^\mu )=1 , \ N_{gh}(C_1)=2.
\]
By following the general procedure we find
\be
\lb{afe}
S =- \int d^4x\   [ \ti{B} (2d\ti{A}  +
\ti{A}\ti{A})  -\fr{1}{2} \ti{A}\ti{A}],
\ee
where the generalized fields are
\[
\begin{array}{ll}
\ti{A} = &  A_{(1+0)} +B^\star_{(2-1)} +C_{0(3-2)}^\star
+C_{1(4-3)}^\star ,\\
\ti{B} = &  B_{(2+0)} -A^\star_{(3-1)} +C_{0(1+1)} +C_{1(0+2)}.
\end{array}
\]
In terms of the components (\ref{afe}) reads
\be
\lb{ath}
H=-\int d^4x\ \{ B_{\mu\nu}F^{\mu\nu} +
2\ep_{\mu \nu \rho \sig }C_0^\mu D^\nu B^{\star \rho \sig}
+2C_1 D^\mu C^\star_{0\mu} +\ep^{\mu \nu \rho \sig }
C_1 [B^\star_{\mu \nu }, B_{\rho \sig}^\star ] -
\fr{1}{2}A_\mu A^\mu \} .
\ee
This is the minimal solution of the
master equation of the theory defined by
(\ref{si})\cite{af}.

\vspace{.1in}

\begin{center}
{\bf \large Acknowledgments}
\end{center}

I would like to thank M. Blau and G. Thompson for
discussions which were essential to complete
this work.

I also thank
Professor Abdus Salam, the International
Atomic Energy Agency and UNESCO for hospitality at the ICTP.

This work is partially supported by the Turkish Scientific and
Technological Research Council (TBTAK).

\pagebreak

\end{document}